\documentclass[twocolumn,twoside,slac_two]{revtex4}
\usepackage{graphicx}
\usepackage{fancyhdr}
\usepackage{hyperref}
\hypersetup{
    colorlinks=true,
    urlcolor=magenta,
}
\begin{document}

\title{Secondary positrons and electrons measured by PAMELA experiment}

%

\author{V.V. Mikhailov$^1$,O. Adriani$^{7}$,
G.  Barbarino$^{2}$, G.A. Bazilevskaya$^{3}$, R. Bellotti$^{4}$, M.
Boezio$^{5}$, E.A.  Bogomolov$^{6}$, M. Bongi$^{7}$, V.
Bonvicini$^{5}$, S. Bottai$^{7}$, A. Bruno$^{4}$, F.S. Cafagna$^{4}$,
D. Campana$^{2}$,  P. Carlson$^{8}$, M.
Casolino$^{9}$, G. Castellini$^{10}$, C. De Donato$^{9}$, C. De
Santis$^{9}$, N. De Simone$^{9}$, V. Di Felice$^{9}$,  A.M.
Galper$^{1}$, A.V. Karelin$^{1}$, S. V. Koldashov$^{1}$, S.
Koldobsky$^{1}$, S.Yu. Krutkov$^{6}$, A.N. Kvashnin$^{3}$, A. A.
Leonov$^{1,3}$,  V.V. Malakhov$^{1}$, Yu.V. Mikhailova$^{1}$,  L. Marcelli$^{9}$, M.
Martucci$^{9,12}$, A.G. Mayorov$^{1}$, W. Menn$^{11}$,M.
Merge$^{9,13}$ E. Mocchiutti$^{5}$, A. Monaco$^{4}$, N.
Mori$^{10}$,R. Munini$^{5}$ G. Osteria$^{2}$, P. Papini$^{7}$, F.
Palma$^{9,13}$, B. Panico$^{2}$, M. Pearce$^{8}$, P.
Picozza$^{9,13}$, M. Ricci$^{12}$, S.B. Ricciarini$^{7}$,  M. F. Runtso$^{1}$, M. Simon$^{11}$,  R. Sparvoli$^{9,13}$,
P. Spillantini$^{1}$, Yu. I. Stozhkov$^{3}$,  A. Vacchi$^{5}$, E.
Vannuccini$^{7}$, G.I. Vasiliev$^{6}$, S.A. Voronov$^{1}$,
 Yu. T. Yurkin$^{1}$, G. Zampa$^{5}$, N. Zampa$^{5}$}
%

\affiliation{$^1$  National Research Nuclear University MEPhI (Moscow Engineering Physics Institute), Kashirskoe highway 31, Moscow, 115409, Russia\\
$^2$INFN, Sezione di Naples and Physics Department of University of Naples Federico II\\
$^3$Lebedev Physical Institute, Russia\\
 $^4$INFN, Sezione di Bari Physics and Department of University of Bari, Italy\\
  $^5$INFN,Sezione di Trieste, Italy\\
$^6$Ioffe Physical Technical Institute,  Russia\\ $^7$INFN,
Sezione di Florence and Physics Department of University of
Florence, Italy\\ $^8$KTH, Department of Physics, and The Oskar
Klein Centre for Cosmoparticle Physics, Sweden\\
 $^9$INFN, Sezione di Rome "Tor Vergata", Italy\\
  $^{10}$INFN, IFAC, Italy\\
$^{11}$Universit\"{a}t Siegen, Department of Physics,Siegen, Germany\\
$^{12}$INFN, Laboratori Nazionali di Frascati, Italy\\
 $^{13}$ University of Rome Tor Vergata, Department of Physics, Italy\\
 $^{14}$Agenzia Spaziale Italiana (ASI) Science Data Center,
Frascati, Italy }

\begin{abstract}
We present a measurements of electron and positron fluxes below the geomagnetic cutoff rigidity in wide energy range from 50 MeV to several GeV by the PAMELA magnetic spectrometer. The instrument was launched on June 15th 2006 on-board the Resurs-DK satellite on low orbit with 70 degrees inclination and altitude between 350 and 600 km. The procedure of trajectories calculations in the geomagnetic field separates stably trapped and albedo components produced in interactions of cosmic ray protons with the residual atmosphere from galactic cosmic rays. Features of spatial distributions of secondary electrons and positrons in the near Earth space, including the South Atlantic Anomaly, were investigated.
\end{abstract}

\maketitle

\thispagestyle{fancy}


High energy primary cosmic rays   produce secondary particles in nuclear interations in atmosphere.  A part  of charged secondaries produced at the top of atmosphere can travel backward in space along the Earth's magnetic field lines.
Flux of secondary electrons with energy E$>$100 MeV  was first calculated in paper \cite{Grigorov}. In this work it was considered process of charged
pions  decay produced  in cosmic ray proton interactions. Pions  decay throw $\pi$ $^{\pm}$ $->$ $\mu$ $^{\pm}$ $->$ $e$ $^{\pm}$ chain to electrons and positrons.
Because production rate is proportional to cosmic ray intensity \emph{I$_{cr}$}, residual atmosphere density $\rho$(\emph{h}) and particle's time of live \emph{T} is $\propto$1/$\rho$(\emph{h}) then resulting  intensity of secondary particles  \emph{J(h)}$\propto$ \emph{I$_{cr}$}$\cdot$ $\rho$ $\cdot$\emph{T}
 will be approximately constant with altitude in wide range forming some kind of a halo around the Earth.
Nuclear interactions of trapped protons of energies $>$300 MeV with upper atmosphere constituents  could be  considered  for
the  production  of  positrons and  electrons  via the same pion decay process in  the innermost  magnetosphere. Calculations performed in paper \cite{Gusev2001} predict the existence of a secondary   belt  in a  narrow  region  around  L-shell  =  1.2. The flux ratio of positrons to electrons e+/e- is
estimated to be $\sim$ 4 for energies of 40-1500 MeV.  This mechanism exhibits sharply
decreasing spectrum with energy. At 200-300 MeV energies, positron fluxes from the trapped proton source are still
comparable with the cosmic ray born positrons, but at higher energies production by trapped protons is negligible.

The magnetic spectrometer PAMELA was launched onboard the
Resurs-DK1 satellite on the 15th of June
2006.  The satellite had a  quasi-polar (70$^{0}$ inclination)
elliptical orbit at an altitudes between 350 and 600 km. Preliminary results of PAMELA observation of secondary electron and positron fluxes near the Earth  made in first year  of the flight were reported in paper \cite{Liuba2009}. The backtracing procedure which determines 
 tracks of particles before their detection offers new
opportunities to compare data with models and it allows to determine
individual particles origin \cite{Plyaskin,Ams01}. This work
presents PAMELA spectrometer measurements of spatial distributions
of secondaries electrons and positrons made using trajectory analysis.


\section{PAMELA spectrometer}
 The instrument consists of a Time-of-Flight system (TOF), an
anticoincidence system, a magnetic spectrometer, an
electromagnetic calorimeter, a shower tail catching scintillator
and a neutron detector . The TOF system provides the main trigger for
particle acquisition, measures the absolute value of the particle
charge and its flight time while crossing the apparatus (the
resolution is better than 350 psec). A rigidity is determined by the
magnetic spectrometer, composed by a permanent magnet with a
magnetic field intensity 0.4 T and a set of six double-side micro-strip
silicon detectors.
The electron and positron identification is provided by the
imaging calorimeter which consist of a series of 44 strip silicon layers interleaved by
22 tungsten planes (16.3 radiation and 0.6 nuclear interaction
lengths deep). Particles not cleanly entering the PAMELA
acceptance are rejected by the anticoincidence system. Using of
the TOF system, the magnetic spectrometer and additional analysis
of the calorimeter information allows extracting electrons and positrons and
measuring their energy (from 50 MeV to several hundred GeV)
effectively. The acceptance of the instrument is about 21.6 cm${^2}$sr
\cite{Picozza2007}.

\section{Data analysis}


 For each registered event the following parameters were measured
or calculated: the number of tracks;  averaged energy losses in the
magnetic spectrometer planes; the rigidity and the track length
(by fitting the track in the magnetic field); the time of flight.
A particle velocity was calculated using the time of flight and
the track length. Moreover, a set of variables dealing with point
of interaction, transversal and longitudinal profiles was
calculated using the calorimeter data \cite{Picozza2007}.

\begin{figure}[h]
\includegraphics[width=20pc]{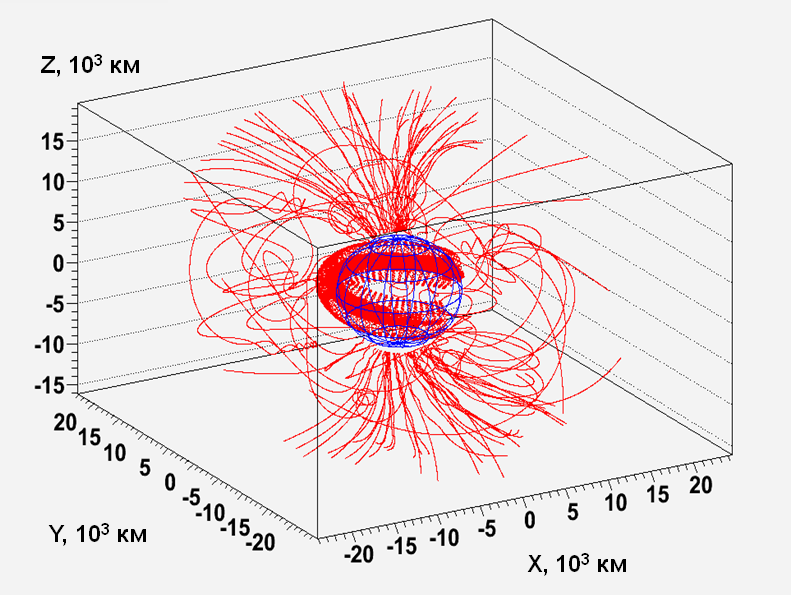}\hspace{2pc}%
\begin{minipage}[b]{14pc}\caption{\label{label} Examples calculated
  trajectories of electrons and positrons detected during several orbits}
\end{minipage}
\end{figure}

Electrons and positrons were identified using information about
\emph{dE/dx} energy losses in the spectrometer planes to determine
charge Z=1 , shower properties in the electromagnetic
calorimeter, particle velocity and a rigidity. The
misidentification of protons and pions is the largest source of
background. Particle identification based on the calorimeter data
can be tuned to rejection power 10$^{4}$ - 10$^{5}$ for protons
and pions, while selecting $>$ 80-90\%  of the electrons or
positrons over all energy range.

Total accumulated statistic for electrons and positron is about
4x10${^6}$ in whole energy range.

Gathering power of the instrument was estimated with Monte-Carlo simulation with official
 PAMELA Collaboration software  \cite{Picozza2007}. An efficiency of the instrument may  change
with time and must be taken into account carefully in a
data proceeding. It was verified from experimental data itself by using different combination of
  information from imaging calorimeter, magnetic spectrometer and  time of flight system.

\begin{figure}[h]
\includegraphics[width=16pc]{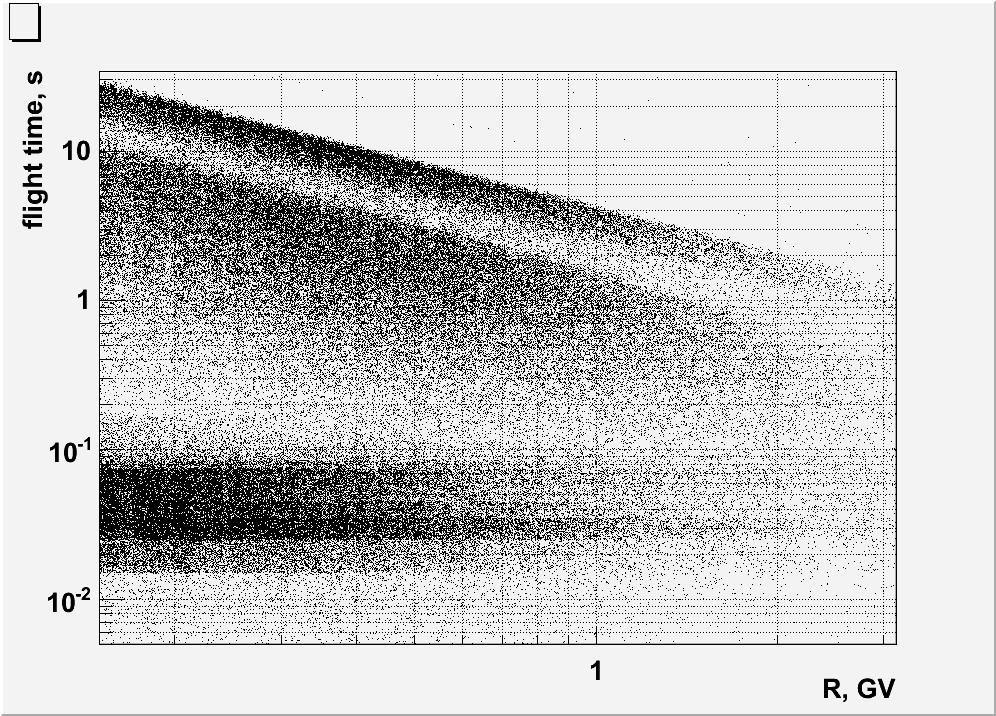}\hspace{2pc}%
\begin{minipage}[b]{14pc}\caption{\label{label}Flight time vs particle's rigidity.}
\end{minipage}
\end{figure}

Using a geographical coordinates and an orientation of PAMELA as a
function of time, the McIlvain geomagnetic coordinates L-shell and
B were calculated for every event. For the calculation, the IGRF
model (http://nssdcftp.gsfc.nasa.gov/models/geomagnetic/igrf) of the
Earth magnetic field was used.

The main axis of PAMELA points to a local zenith. Orbit
characteristics allow measuring particles with pitch angles (the
angle between the particle velocity and the magnetic field vector)
of about 80-90$^{0}$ in the equatorial region. 


Primary particles are observed mainly in polar regions and above cut-off rigidity
 near equator. This component consist of mainly electrons. Secondary 
component with small pitch-angles  is observed below geomagnetic cut-off  for all latitudies and secondary trapped particles can be measured in South Atlantic Anomaly (SAA) with pitch-angles about 90 degree \cite{Liuba2009, Adriani2013}.

To determine primary or secondary origin of detected particles  the  backtracing procedure was
applied for every identified events to obtain trajectory of events
up to 35 second before detection.  The tracing  was stopped if particles
touched the Earth atmosphere on 40 km altitude or escaped
magnetosphere. Escaping boundary  was chosen to be 20000 km.
Trajectory allows to determine particle origin for every individual
event.   Figure 1 shows examples calculated
  trajectories of electrons and positrons detected during several orbits. Totally  about $~$10 $^{6}$ events were backtraced.  Finally,  primary electrons and positrons with high rigidity R were excluded from analysis together with particles in transition region near geomagnetic cut-off , applying condition R$<$ 10/L$^{3}$ [GV],  where L is geomagnetic L-shell.


\begin{figure}[h]
\begin{minipage}{14pc}
\includegraphics[width=14pc]{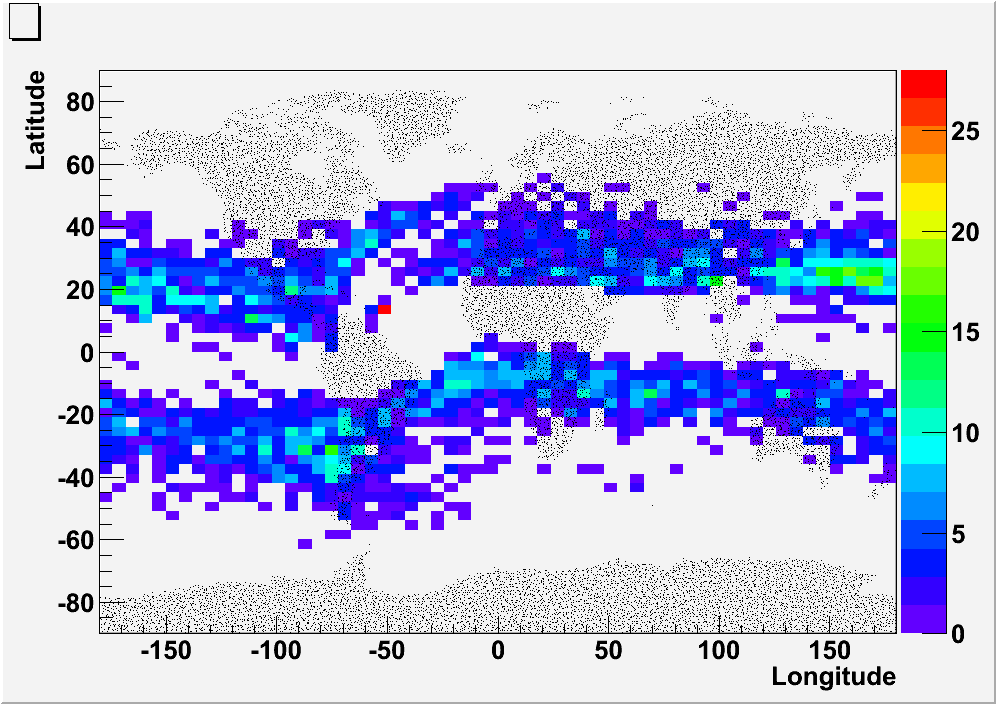}
\caption{\label{label}Regions where short-lived positrons were
originated.}
\end{minipage}\hspace{2pc}%
\begin{minipage}{14pc}
\includegraphics[width=14pc]{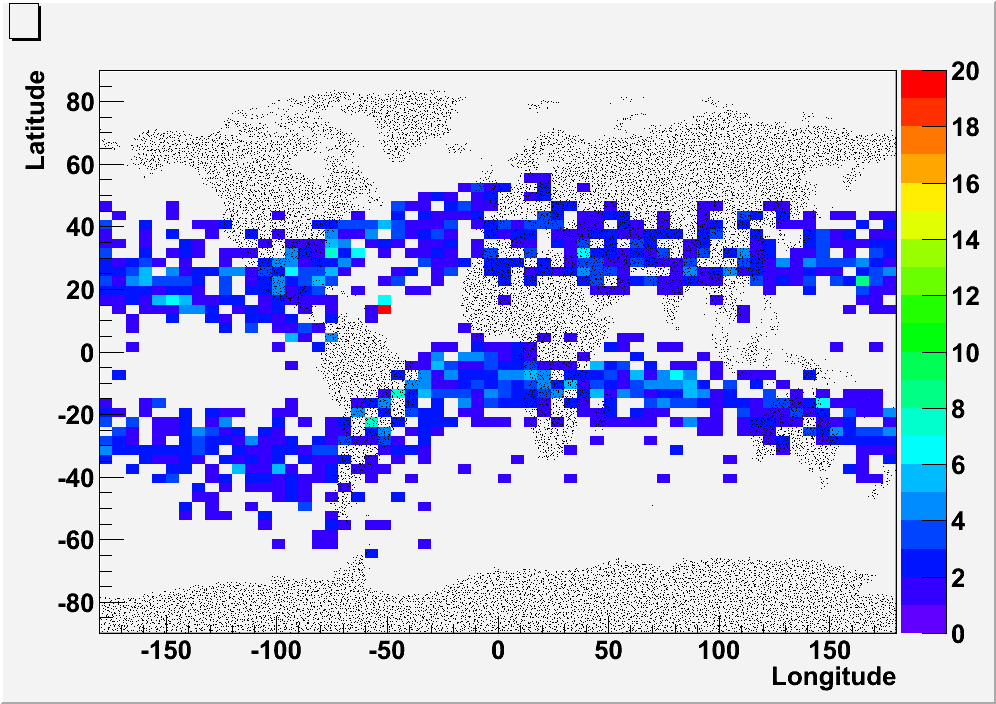}
\caption{\label{label}Regions where short-lived electrons were
originated.}
\end{minipage}
\end{figure}


\begin{figure}[h]
\begin{minipage}{14pc}
\includegraphics[width=14pc]{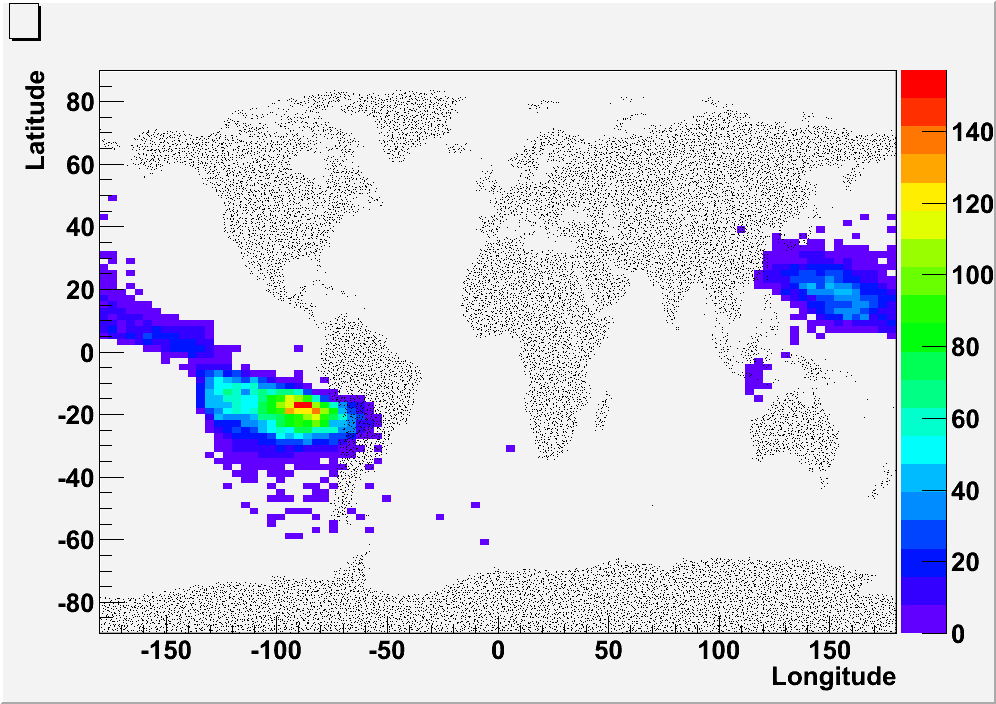}
\caption{\label{label}Regions where quasi-trapped positrons were
originated.}
\end{minipage}\hspace{2pc}%
\begin{minipage}{14pc}
\includegraphics[width=14pc]{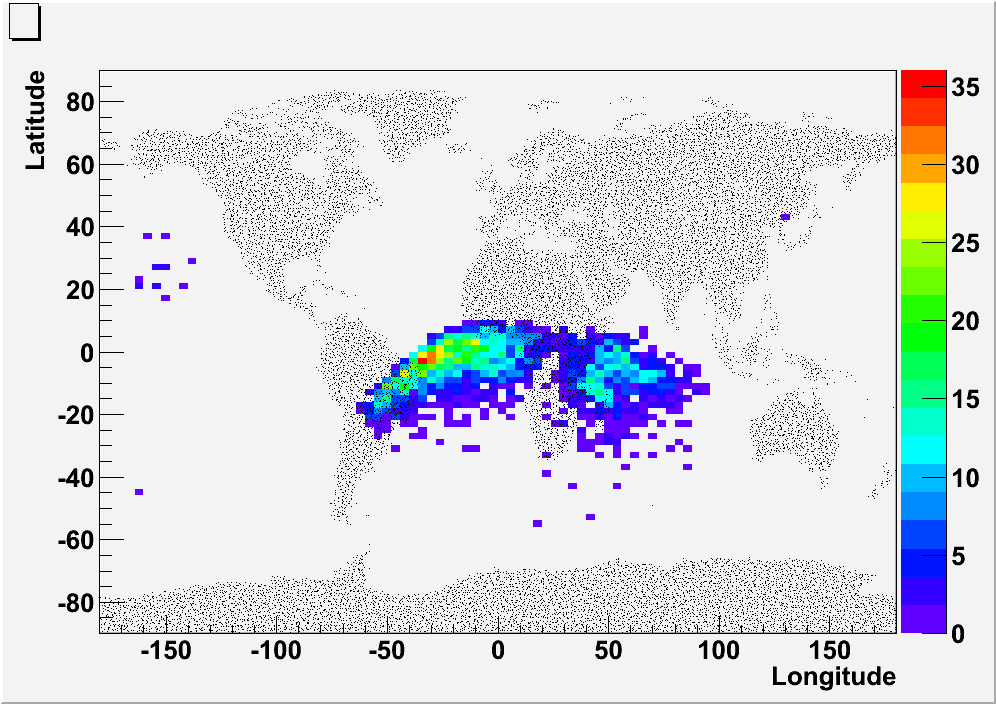}
\caption{\label{label}Regions where quasi-trapped electrons were
originated.}
\end{minipage}
\end{figure}

\begin{figure}[h]
\begin{minipage}{14pc}
\includegraphics[width=14pc]{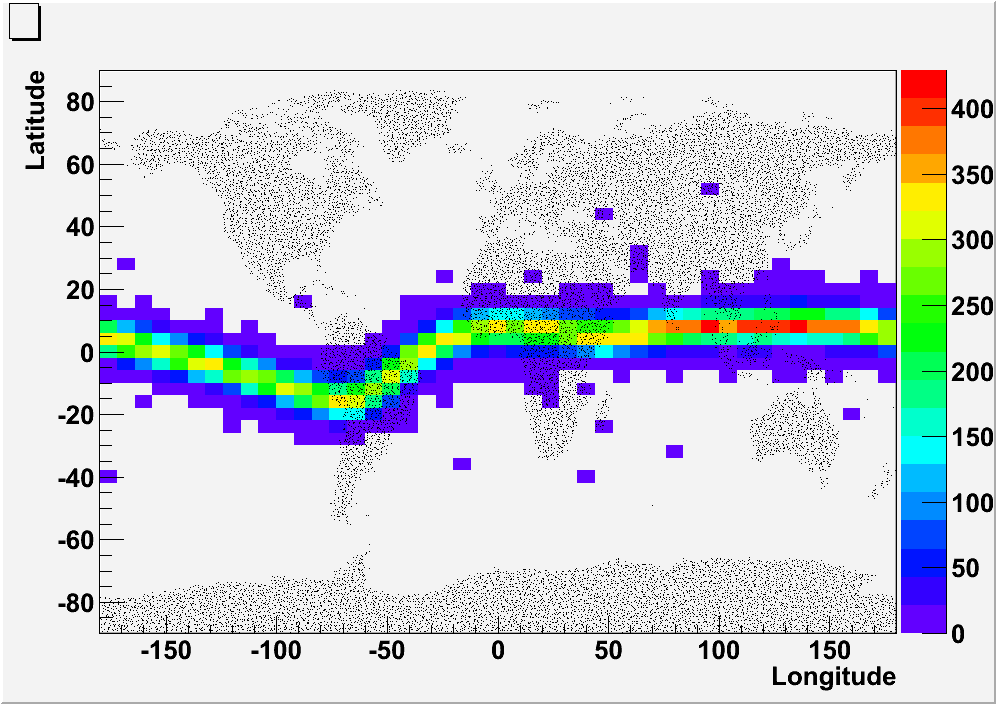}
\caption{\label{label}Regions where trapped positrons were
35 second before detection.}
\end{minipage}\hspace{2pc}%
\begin{minipage}{14pc}
\includegraphics[width=14pc]{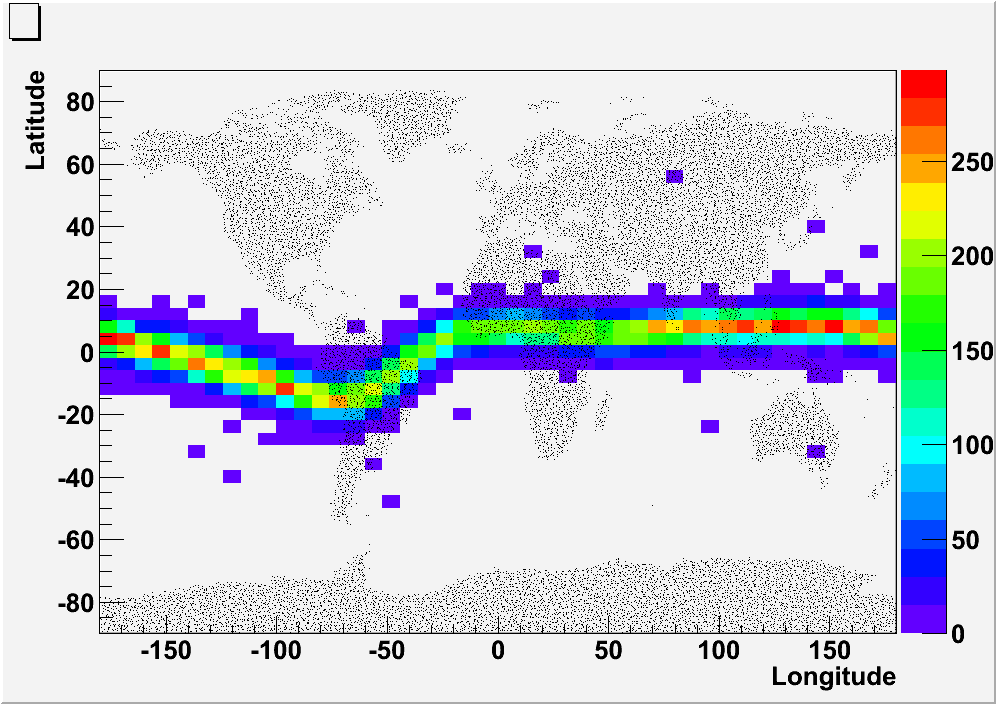}
\caption{\label{label}Regions where trapped electrons were
35 second before detection.}
\end{minipage}
\end{figure}

  \section{Results}
Calculations of particle trajectories in geomagnetic field provides additional information about
spatial distribution of secondary fluxes in the Earth vicinity because it gives possibility to explore fluxes in enlarged regions outside the satellite orbits.
In figure 2 time of flight in magnetosphere versus of rigidity in magnetosphere is shown for equatorial region (L $<$2).
Typically reentrant albedo particles reach the orbit of the satellite for time  about  0.1 second. For relativistic electrons and positrons this flight time does not dependent from particle's rigidity.   But if electron or positron has appropriate pitch-angle to be mirrored by magnetic field , flight time iincreases dramatically due to drift around the Earth. Drift speed is increasing with  increasing of particle rigidity and   so  flight time of   quasi-trapped particles is decreasing with rigidity as it is seen from figure.
Such behavior of secondary albedo and quasi-trapped electrons and positrons correspond to that  was observed in AMS-01 experiment \cite{Ams01} where they were called as short-lived and long-lived second leptons.

In figure 3,4  points of origin  of reentrant  electrons
and positrons are shown.,
With agreement with previous observations by AMS-01 quasi-trapped electrons and positrons originated from specific regions (figure 5,6).

There is also a significant part of track
with very large live-time which can not be determined by simple tracing
procedure. Figure 7,8  shows calculated spatial position
 of  trapped electrons and positrons they had 35 second before detection.

Figure 9 shows positron to electron ratio  for different components.  Results for quasi-trapped and short-lived albedo particles are consistent with measurements made in AMS-01 experiment \cite{Ams01} shown by open points in the figure. The ratio of positrons and electrons fluxes for stably trapped particles is much lower then for quasi-trapped particles. The maximum counting rate of the trapped particles in the experiment observed on L-shell 1.18 and increases with decreasing in the value of the magnetic field. These results are difficult to explain in terms of the model of inelastic interaction of radiation belt protons with residual atmosphere \cite{Gusev2001}, which predicts significant  increase in  positrons at least at low energies. Results of measurements   suggest a more efficient capture of electrons due to different boundary conditions for electrons and positrons near SAA  or different time of live of trapped positrons and electrons.

\begin{figure}[h]
\includegraphics[width=20pc]{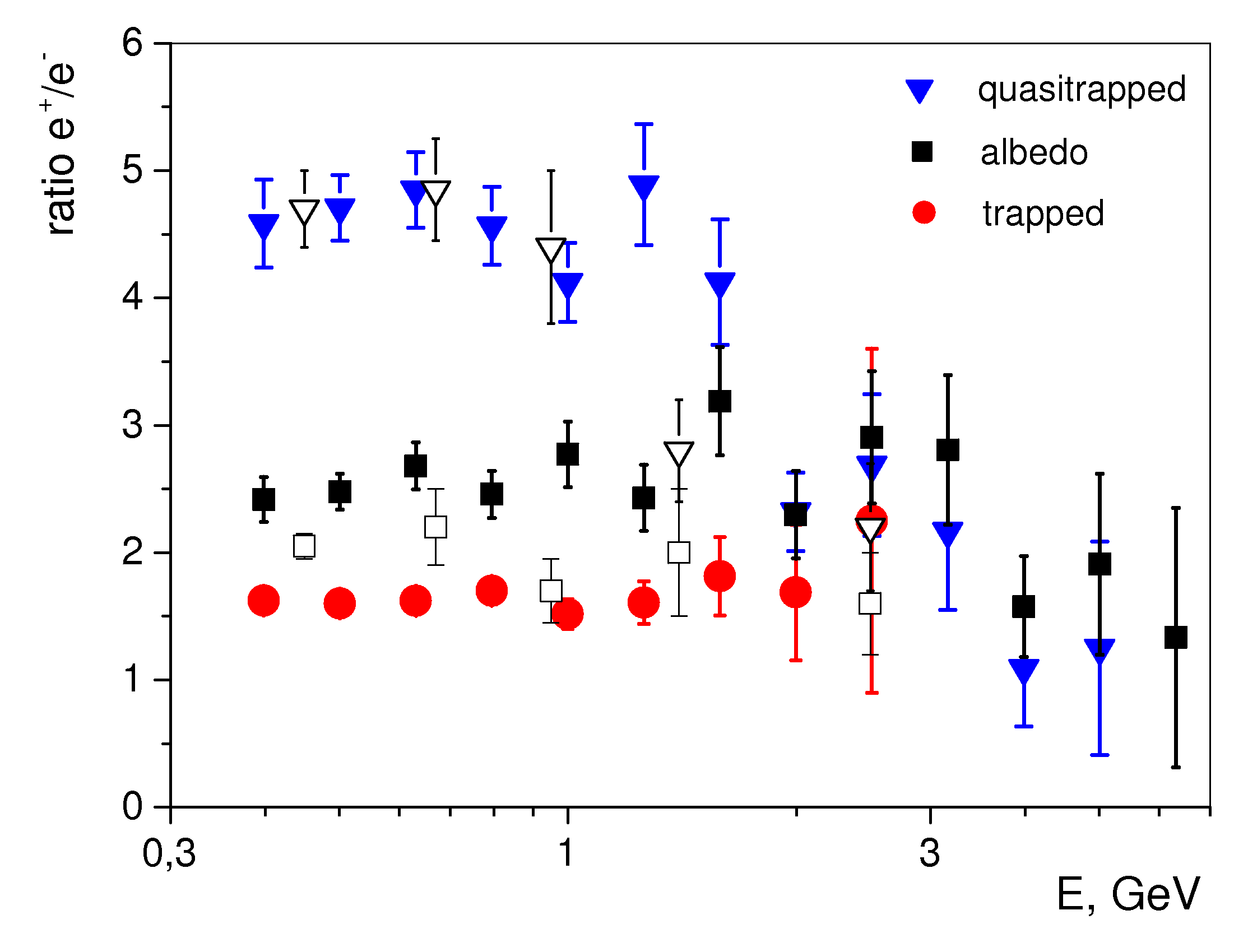}\hspace{2pc}%
\begin{minipage}[b]{14pc}\caption{\label{label}e$^{+}$/e$^{-}$  ratio
 vs energy for quasi-trapped, albedo and trapped components. AMS-01 data are shown by open points.}
\end{minipage}
\end{figure}

\section{Summary}

Secondary electron and positron fluxes have complex spatial
structure caused by geomagnetic field,  production cross-section
and atmospheric absorption of produced secondaries. Using backtracking procedure over ~10$^{6}$ events three components of secondary positrons were clearly separated: albedo, quasi-trapped and stably trapped.   Measured positron to electron ratio in
the PAMELA experiment points out on different production mechanism of trapped and quasi-trapped particles .

\section{Acknowlegments}
We acknowledge support from Russian Federal Space Agency (Roscosmos), the Italian Space Agency (ASI), Deutsches Zentrum fur Luft- und Raumfahrt (DLR), the Swedish National Space Board. V.V. M. would like  to thank for the support from National Research Nuclear University MEPhI in the framework of the Russian Academic Excellence Project (contract No. 02.a03.21.0005, 27.08.2013).

\bigskip 

\end{document}